\documentclass[aip,reprint,pop]{revtex4-1}

\usepackage{graphicx}
\usepackage{subfigure}
\usepackage{epstopdf}
\usepackage{amsmath}
\usepackage{nicefrac}
\usepackage{dcolumn}
\usepackage{bm}
\usepackage{mathrsfs}
\usepackage[colorlinks=true, linkcolor=blue]{hyperref}

\begin{document}

\title{Nonlinear Ohmic Dissipation in Axisymmetric DC and RF Driven Rotating Plasmas}
\author{J.M. Rax}
\affiliation{Departments of Physics, University of Paris XI \& Ecole
Polytechnique, LOA, 91128 Palaiseau France}
\author{E. J. Kolmes, I. E. Ochs, N. J. Fisch}
\affiliation{Princeton Plasmas Physics Laboratory, Princeton University,
Princeton, New Jersey, 08540 USA}
\author{R. Gueroult}
\affiliation{LAPLACE, Universit\'{e} de Toulouse, CNRS, 31062 Toulouse, France}

\begin{abstract}
An axisymmetric fully ionized plasma rotates around its axis when a charge separation between magnetic
surfaces is produced from DC fields or RF waves. On each magnetic surface both electrons and ions obey the isorotation law and perform an azimuthal E cross B rotation at the same
angular velocity. When Coulomb collisions are taken into account such a flow
displays no Ohmic current short circuiting of the charge separation and thus no
linear dissipation. A nonlinear Ohmic response appears when inertial effects
are considered, providing a dissipative relaxation of the charge separation
between the magnetic surfaces. This nonlinear conductivity results from an
interplay between Coriolis, centrifugal and electron-ion collisional
friction forces. This phenomena is identified, described and analyzed. In addition, both the quality factor of angular momentum storage as well as the efficiency of
wave driven angular momentum generation are calculated and shown to be
independent of the details of the charge separation processes.
\end{abstract}

\date{\today}
\maketitle

\section{Introduction}
\label{Sec:Sec1}

Axisymmetric, fully ionized, magnetized plasmas can be set in rotation
around their axis through a small breakdown of quasineutrality sustained
by DC or RF power. The uncompensated polarizing free charges rapidly
rearrange along the magnetic field lines to screen the electric field
component along the magnetic field. A steady state electric field $\mathbf{E}
$, perpendicular to the magnetic field $\mathbf{B}$, drives an $E$ cross $B$
azimuthal flow around the axis~\cite{Lehnert1971,Wilcox1959}. The steady-state sustainment of the
charge imbalance between magnetic surfaces can be achieved either (\textit{i)%
} with an applied DC radial voltage drop between magnetic surfaces or (%
\textit{ii}) through resonant wave induced radial drift across magnetic
surfaces. The first scheme requires a set of polarized end plates
intercepting the magnetic field lines at the edge of the discharge, the
second scheme needs the propagation and absorption of a plasma wave with the
right dispersion relation and carrying a significant amount of angular
momentum around the axis. Both schemes are associated with injected power
consumption. This power is ultimately dissipated through the nonlinear Ohmic
radial current identified and described in this study.

Two classical rotating configurations have been widely investigated in
plasma physics : the \textit{Brillouin rigid body rotation} associated with
homogeneous magnetic fields~\cite{Brillouin1945}, and the \textit{Ferraro isorotation }
associated with an axisymmetric inhomogeneous magnetic fields~\cite{Ferraro1937}. The\
study of axisymmetric rotating pulsar magnetospheres offers a third~\cite{Beskin1986,Beskin1997},
more complex, model of rotating configurations where the uncompensated
charges density $\rho $ is called the \textit{Goldreich-Julian charge
density }(G-J)~\cite{Goldreich1969,Julian1973}. Because of the absence of relativistic and radiative
effects, the axisymmetric laboratory plasmas analyzed in this paper are
much simpler than pulsar magnetospheres. Nevertheless we will adopt this
nomenclature and call \textit{G-J charges} the uncompensated charges driving
the rotation to mark the differences with the background quasineutral
charges. Beside pulsar magnetospheres, rotating axisymmetric plasmas and the
general problem of angular momentum conversion and dissipation with radial
electric field, axial magnetic field, wave helicity and plasma vorticity,
has received considerable attention within the framework of : (\textit{i})
plasma centrifuge for isotope separation~\cite{Bonnevier1966,Ohkawa2002,Shinohara2007,Gueroult2014,Fetterman2011,Prasad1987}, (\textit{ii}) nonneutral
plasma physics~\cite{Davidson2001,Davidson1969}, (\textit{iii}) thermonuclear magnetic confinement
studies with homopolar devices, rotating mirrors and tokamaks~\cite{Fetterman2008,Bekhtenev1980,Fetterman2011a,Rax2017}, (%
\textit{iv)} particle acceleration and magnetic field generation with plasma
bubbles and channels~\cite{Shvets2002,Kostyukov2002,Thaury2013}.

If we consider the classical MHD Ohm's law on an axisymmetric magnetic
surface rotating at velocity $\mathbf{v}=$ $\mathbf{E\times B/}B^{2}$, such
that $\mathbf{E}\cdot \mathbf{B}=0$, then the current 
\begin{equation}
\mathbf{j}=\sigma \left( \mathbf{E}+\mathbf{v}\times \mathbf{B}\right) \text{%
,}  \label{ohoh}
\end{equation}
where $\sigma $ is the conductivity~\cite{Spitzer1952}, cancels. This cancellation of
the linear Ohmic response is exact only if the $E$ cross $B$ drift is
uniform~\cite{Rozhansky2008}. If the $E$ cross $B$ drift is accelerated Eq.~(\ref{ohoh})
must be complemented with inertial forces and, as we will demonstrate, a
nonlinear Ohmic response comes into play to\ ensure G-J charges relaxation
and dissipation.

If we study a conducting fluid of free particles with charge $q$ and mass $%
m$ per particle, rather than Eq.~(\ref{ohoh}) we must consider the relation~\cite{Landau1984}
\begin{equation}
\mathbf{j}=\sigma \left( \mathbf{E}+\mathbf{v}\times \mathbf{B-}\frac{m}{q}%
\mathbf{v}\cdot \mathbf{\nabla v}\right) \text{.}  \label{ohoh2}
\end{equation}
For electrons in metals the effects associated with the last term of the
right hand side of Eq.~(\ref{ohoh2}) are called \textit{excitation of
current by acceleration} and Eq.~(\ref{ohoh2}) provides the right tool to
describe rotating metallic conductors~\cite{Landau1984}.

For ions and electrons in fully ionized plasmas these effects are called 
\textit{inertial effects} and Eq.~(\ref{ohoh2}) must be revisited with a two
fluid model~\cite{Rozhansky2008,Helander2002}. Inertial effects, associated with centrifugal and Coriolis forces, are
usually small, but, as we will demonstrate, they must be taken into account
to describe the collisional Ohmic short circuiting of the G-J charges both
for Brillouin and Ferraro flows.

Beside inertial effects, \textit{finite Larmor radius effects}, driving
dissipative diamagnetic flows, and \textit{ion-ion collisions}, driving
viscous momentum transfer~\cite{Rozhansky2008,Helander2002}, are also responsible of a small
dissipative current short circuiting the G-J charges. For plasmas considered
in this study the order of magnitude of these finite Larmor radius and
viscous effects is evaluated and shown to be typically smaller than inertial
effects for Brillouin and quasi-Brillouin flows.

To analyse the impact of the inertial effects, rather than the one fluid MHD
relation Eq.~(\ref{ohoh2}), we will use a two fluids model all along this
study. This paper is organized as follows. In Sec.~\ref{Sec:Sec2} we consider a rigid
body, fully ionized, collisional Brillouin flow and set up the two coupled,
fourth order, algebraic equations fulfilled by the ion and electron
vorticities. These equations are solved for the slow branch of Brillouin
rotation modes through an expansion with respect to collisionality and a
dissipative nonlinear current is found to provide G-J charge relaxation. The
conductivity is nonlinear and displays a quadratic scaling with respect to
the electric field. The origin of this conduction is rather intricate as it
involves an interplay between Coriolis, centrifugal and Coulomb friction
forces. A step by step physical analysis of this response is proposed in
Sec.~\ref{Sec:Sec4}. In Sec.~\ref{Sec:Sec3}, to prepare the study of Sec.~\ref{Sec:Sec4}, we review the basic
electrodynamics of rotating axisymmetric plasmas. We derive the Ferraro
isorotation law as well as some new expressions for the G-J charges.

In Sec.~\ref{Sec:Sec4} we calculate the conductivity induced by electron-ion collisions
combined with inertial drifts for a fully ionized discharge fulfilling
isorotation in an axisymmetric magnetic configuration. Then, in Sec.~\ref{Sec:Sec5}, this
new result is used (\textit{i}) to evaluate the \textit{quality factor} of
energy storage in a rotating plasma and, in Sec.~\ref{Sec:Sec6}, (\textit{ii}) to
describe and analyse the efficiency of wave driven rotation. We calculate the efficiency of wave orbital angular momentum conversion into
plasma orbital angular momentum and show that it is independent of the
details of the wave dispersion and the wave-particle resonance. These new
results on the quality factor and the efficiency clearly displays the
interest of schemes with wave sustainment of plasma rotation which have
been put forward both for isotope separation and magnetic confinement~\cite{Fetterman2011,Fetterman2008,Rax2017}.

The impact of diamagnetic flows is then analyzed in Sec.~\ref{Sec:Sec7}. This finite
Larmor radius effect induces a linear electric conduction and this current
is shown to be smaller than the inertial nonlinear current. Section~\ref{Sec:Sec7} is
also devoted to the comparison between viscous damping, resulting from the
coupling between slow and fast magnetic surfaces, and Ohmic dissipation. The
original results of this study are summarized in Sec.~\ref{Sec:Sec8} concluding this
study.

\section{Nonlinear Ohmic conduction in a Brillouin flow}
\label{Sec:Sec2}

In this section we set up the two coupled nonlinear algebraic equations,
Eqs.~(\ref{alg1}) and (\ref{alg2}), fulfilled by electrons and ions
vorticities describing a collisional, fully ionized, Brillouin flow~\cite{Brillouin1945,Davidson2001}.
Then, we expand the solutions of these equations with respect to the
collision frequency and express the nonlinear conductivity resulting from
this expansion. The extension and confirmation of the final result of this
section to a general axisymmetric flows is considered and analyzed in the
following sections.

\begin{figure*}
\begin{center}
\includegraphics[width = 10cm]{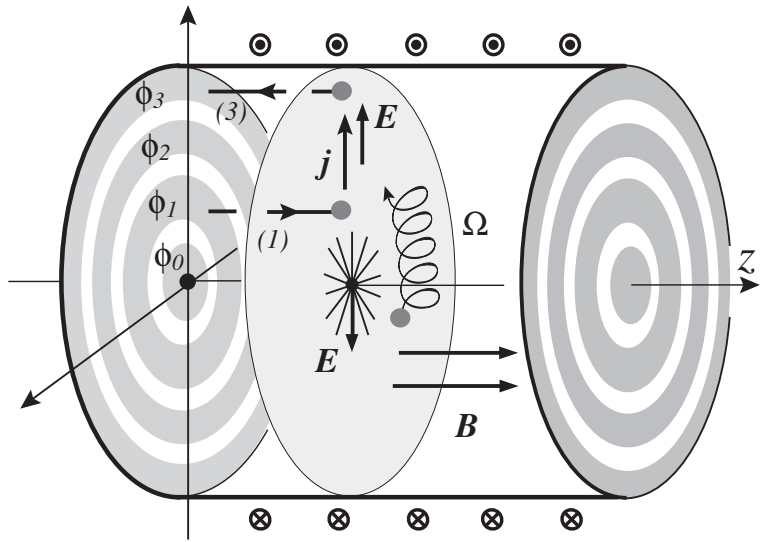}
\caption{Fully ionized Brillouin flow driven by a set of concentric
electrodes sustaining a radial electric field $\mathbf{E}$, the current path
from $\phi _{1}$ to $\phi _{3}$ is made of two highly conducting axial field
lines $(1)$ and $\left( 3\right) $ short circuited by the radial nonlinear
Ohmic current $\mathbf{j}$. }
\label{Fig:Fig1}
\end{center}
\end{figure*}

Consider a rotating, cylindrical, fully ionized, uniform, magnetized plasma
illustrated in Fig.~\ref{Fig:Fig1} where a set of concentric polarized electrodes
provides a radial voltage drop ($\phi _{0}>\phi _{1}>\phi _{2}...$) between
the various cylindrical magnetic surfaces. We use cylindrical polar
coordinates $\left( r,\theta ,z\right) $ associated with the cylindrical
polar basis $\left[ \mathbf{e}_{r},\mathbf{e}_{\theta },\mathbf{e}%
_{z}\right] $. The magnetic field $\mathbf{B}$ is uniform and directed along
the $z$ axis and the electric field $\mathbf{E}$ is radial and increases
linearly with respect to $r$%
\begin{equation}
\mathbf{B}=B\mathbf{e}_{z}\text{ , }\mathbf{E}=E\left( r\right) \mathbf{e}%
_{r}\text{.}
\end{equation}
As $\mathbf{\nabla }\cdot \mathbf{B}=0$ we can introduce the flux function $%
2\pi \Psi \left( r\right) $ to describe the magnetic field. As $\mathbf{%
\nabla }\times \mathbf{E}=\mathbf{0}$ we can introduce the electrostatic
potential $\phi \left( r\right) $ to describe the electric field. With these
flux and potential functions the expressions of the fields becomes
\begin{equation}
\mathbf{B}=\mathbf{\nabla }\Psi \times \frac{\mathbf{e}_{\theta }}{r}\text{
, }\mathbf{E}=-\mathbf{\nabla }\phi \text{.}
\end{equation}
Introducing the low frequency permittivity, $\varepsilon _{\bot }$ = $1+{%
\omega }_{pi}^{2}/{\omega }_{i}^{2}$ $\approx {\omega }_{pi}^{2}/{\omega }%
_{i}^{2}$ ($\omega _{pi}$ $=$ $nq^{2}/\varepsilon _{0}m_{i}$ and $\omega
_{i} $ $=$ $qB/m_{i}$ are the ion plasma and ion cyclotron frequencies)
these flux and potential functions can be expressed as 
\begin{equation}
\Psi =B\frac{r^{2}}{2}\text{, }\phi =-\frac{\rho }{\varepsilon
_{0}\varepsilon _{\bot }}\frac{r^{2}}{4}\text{.}
\end{equation}
where $\rho $ is the uniform G-J charges density such that $\mathbf{\nabla }%
\cdot \varepsilon _{\bot }\varepsilon _{0}\mathbf{E}=\rho $. We assume that
heating and fuelling systems, such as waves and pellet injections,
complemented by a particle and power exhaust system, ensure a steady state
complete ionization and a steady state flat density profile. The small
breakdown of quasineutrality $\rho $ is short circuited by a collisional
current $\mathbf{j}$ and is sustained either by wave absorption or DC radial
polarization. For this latter case, a set of concentric electrodes,
illustrated in Fig.~\ref{Fig:Fig1}, provides a voltage drop ($\phi _{0}>\phi _{1}>\phi
_{2}>\phi _{3}$) between the various cylindrical magnetic surfaces as the
conductivity along the field lines is very large.

The collisional depletion of $\rho $ through $\mathbf{j}$ is continuously
compensated by the power supply driving these electrodes, or through wave
induced charge separation, in order to ensure steady state rotation. In this
section, we will demonstrate that $\ \mathbf{j}$ \ is directed along $\ 
\mathbf{E}$, but is not a linear function of $\mathbf{E}$.

In a fully ionized plasma, with electron velocity $\mathbf{v}_{e}$ and ion
velocity $\mathbf{v}_{i}$, electron-ion collisions are the source of
friction forces, $\mathbf{F}_{i\rightarrow e}$ $=-$ $\mathbf{F}%
_{e\rightarrow i}$, between electrons and ions. At the single particle level
these dissipative forces can be expressed as 
\begin{equation}
\mathbf{F}_{i\rightarrow e}=-m_{e}\nu _{e}\left( \mathbf{v}_{e}-\mathbf{v}%
_{i}\right) \text{, }\mathbf{F}_{e\rightarrow i}=-m_{i}\nu _{i}\left( 
\mathbf{v}_{i}-\mathbf{v}_{e}\right),
\end{equation}
where we have introduced $\nu _{e}$ and $\nu _{i}$, the momentum exchange
frequencies between the two populations, according to the classical
definitions 
\begin{equation}
m_{e}\nu _{e}=m_{i}\nu _{i}=\sqrt{\frac{m_{i}m_{e}}{m_{e}+m_{i}}}\frac{%
nq^{4}\log \Lambda }{6\pi {\varepsilon _{0}}^{2}\left( 2\pi \right) ^{\frac{1}{%
2}}\left( k_{B}T\right) ^{\frac{3}{2}}}=m\nu \text{,}
\end{equation}
where $-q$ is the electron charge, $m_{e}$ and $m_{i}$ the electron and ion
masses and the other notations are standard~\cite{Helander2002}. We have considered a fully
ionized hydrogen plasma with a quasineutral density $n$ such that $\rho \ll
nq$. The temperature $T=T_{e}=T_{i}$ is uniform and the effective mass is
defined as $m=\sqrt{m_{i}m_{e}}$.

The relation $m_{e}\nu _{e}=m_{i}\nu _{i}$ describes the fact that the
momentum lost by one population is gained by the other, strictly speaking
the two friction forces are not \textit{dissipative} as there is no \textit{%
entropy production} because the macroscopic momentum is not dispersed into a
large number of microscopic degrees of freedom but transferred, at the
macroscopic level, from the slow population to the fast one if $\mathbf{v}%
_{e}\neq \mathbf{v}_{i}$. The steady-state momentum balance for the
electrons and ions populations is given by two Euler fluid equations coupled
through these Coulomb friction terms $\mathbf{F}_{i\rightarrow e}$ $=-$ $%
\mathbf{F}_{e\rightarrow i}$.

In this section we will follow the method used previously to study the
weakly ionized collisional Brillouin flow~\cite{Rax2015}. The inertial term of Euler's equations, $d\mathbf{v}/dt=\left( \mathbf{v}%
\cdot \mathbf{\nabla }\right) \mathbf{v}$, is expressed through the
classical identity: $\left( \mathbf{v}\cdot \mathbf{\nabla }\right) \mathbf{v%
}$ = $\mathbf{\nabla v}^{2}/2$ + $\left( \mathbf{\nabla }\times \mathbf{v}%
\right) \times \mathbf{v}$. In writing Euler's equations we explicitly
display the axial vorticities $\Omega _{e/i}$ associated with a velocity
field $\mathbf{v}_{e/i}$ and defined as: $\ \Omega _{e/i\text{ }}$ = $%
\left( \mathbf{\nabla }\times \mathbf{v}_{e/i}\right) \cdot \mathbf{b}/2$,
where $\mathbf{b}=\mathbf{B}/B$. The steady state momentum balances between
centrifugal ($\mathbf{\nabla v}^{2}/2$) and Coriolis ($2\Omega \mathbf{b}%
\times \mathbf{v}$) inertial forces, on the left hand side of Eqs.~(\ref{jj1}%
) and (\ref{jjjj}), and electric Coulomb forces, magnetic Laplace forces and
electron-ion friction forces, on the right hand side of Eqs.~(\ref{jj1})
and (\ref{jjjj}), is given by
\begin{subequations}
\begin{multline}
\frac{1}{2}\mathbf{\nabla}{\mathbf{v}_{e}}^{2}+2\Omega _{e}\mathbf{b}\times \mathbf{v}_{e} =-\frac{q}{m_{e}}\mathbf{E}-\frac{q}{m_{e}}\mathbf{v}_{e}\times B\mathbf{b} \\- \nu _{e}\left( \mathbf{v}_{e}-\mathbf{v}_{i}\right),
\label{jj1}
\end{multline}
\begin{multline}
\frac{1}{2}\mathbf{\nabla}{\mathbf{v}_{i}}^{2}+2\Omega _{i}\mathbf{b}\times \mathbf{v}_{i} =\frac{q}{m_{i}}\mathbf{E}+\frac{q}{m_{i}}\mathbf{v}_{i}\times B\mathbf{b} \\- \nu _{i}\left( \mathbf{v}_{i}-\mathbf{v}_{e}\right).
\label{jjjj}
\end{multline}
\end{subequations}

\begin{figure}
\begin{center}
\includegraphics[width = 8cm]{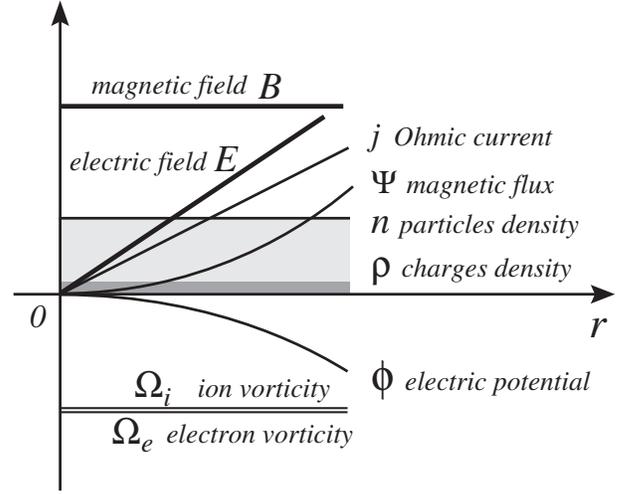}
\caption{Electric potential $\phi $, magnetic flux $\Psi $, vorticity $%
\Omega $, charges density $\rho $ and current $j$ profiles fo a Brillouin
flow. }
\label{Fig:Fig2}
\end{center}
\end{figure}

We do not consider viscous dissipation because $\mathbf{\Delta v}_{e}=%
\mathbf{\Delta v}_{i}=\mathbf{0}$ are exactly fulfilled by the rigid body
rotation solutions. For a generic rotation considered here, if the
temperature, density and vorticity distributions are rather flat with
respect to the radial variable, viscosity effects and diamagnetic responses
are pushed toward the boundary of the plasma and are negligible in the core
of the column, at this boundary the matching of the rotation to the wall is
another problem which requires the identification and analysis of an
electrical, thermal and mechanical boundary layer, this problem is out of
the scope of this study. The typical radial profiles of the various
dynamical quantities (flux, potential, vorticity...) describing this flow
are illustrated in Fig.~\ref{Fig:Fig2}. Euler's equations Eqs.~(\ref{jj1}) and (\ref{jjjj}%
) can be rewritten as
\begin{subequations}
\begin{eqnarray}
&&\mathbf{v}_{e}+\alpha _{e}\mathbf{v}_{e}\times \mathbf{b}=\mathbf{v}_{i}-%
\frac{1}{m\nu }\mathbf{\nabla }\left( m_{e}{v_{e}}^{2}/2-q\phi \right),
\label{e1} \\
&&\mathbf{v}_{i}+\alpha _{i}\mathbf{v}_{i}\times \mathbf{b}=\mathbf{v}_{e}-%
\frac{1}{m\nu }\mathbf{\nabla }\left( m_{i}{v_{i}}^{2}/2+q\phi \right),
\label{e2}
\end{eqnarray}
\end{subequations}
where we have introduced the electron cyclotron frequency $\omega
_{e}=qB/m_{e}$, the ion cyclotron frequency $\omega _{i}$ and defined the
generalized Hall parameters, $\alpha _{e}$ and $\alpha _{i}$,$^{28}$ \
according to the definitions 
\begin{equation}
\nu _{e}\alpha _{e}=\omega _{e}-2\Omega _{e}\text{ , }\nu _{i}\alpha
_{i}=-\omega _{i}-2\Omega _{i}\text{.}
\end{equation}

In order to find the rigid body rotation solutions of equations Eqs.~(\ref
{e1}) and (\ref{e2}), we take (\textit{i}) the rotational and then (\textit{%
ii}) the divergence of these two equations. We consider that all the
variations of the $\mathbf{v}_{e/i}$ and $\phi $ fields are radial so that $%
\mathbf{\nabla }\times \left( \mathbf{v}\times \mathbf{b}\right) $ = $%
-\left( \mathbf{\nabla }\cdot \mathbf{v}\right) \mathbf{b}$ and $\mathbf{%
\nabla }\cdot \left( \mathbf{v}\times \mathbf{b}\right) $ = $\left( \mathbf{%
\nabla }\times \mathbf{v}\right) \cdot \mathbf{b}$. With the help of these
relations we obtain 
\begin{equation}
2\Omega _{e}-\alpha _{e}\mathbf{\nabla }\cdot \mathbf{v}_{e}=2\Omega _{i}%
\text{ , }2\Omega _{i}-\alpha _{i}\mathbf{\nabla }\cdot \mathbf{v}%
_{i}=2\Omega _{e}\text{,}  \label{fji}
\end{equation}
by taking the rotational of Eqs.~(\ref{e1}) and (\ref{e2}). Then, taking the
divergence of Eqs.~(\ref{e1}) and (\ref{e2}) leads to the equations
fulfilled by the vorticities $\Omega _{e/i}$ 
\begin{subequations}
\begin{multline}
\left( \Omega _{e}-\Omega _{i}\right) \left( \frac{1}{\alpha _{e}}+\frac{1}{\alpha _{i}}\right) +\alpha _{e}\Omega _{e} \\ = -\frac{1}{2m\nu }\Delta
\left( m_{e}{v_{e}}^{2}/2-q\phi \right),
\end{multline}
\begin{multline}
\left( \Omega _{i}-\Omega _{e}\right) \left( \frac{1}{\alpha _{e}}+\frac{1}{\alpha _{i}}\right) +\alpha _{i}\Omega _{i} \\ = -\frac{1}{2m\nu }\Delta
\left( m_{i}{v_{i}}^{2}/2+q\phi \right).
\end{multline}
\end{subequations}

Under the rigid body rotation hypothesis~\cite{Rax2015}, the ions and electrons
azimuthal velocities are given by $v_{\theta e}=\Omega _{e}r$ and $v_{\theta
i}=\Omega _{i}r$ and the relations Eq.~(\ref{fji}) give the following radial
components after integration, $v_{re}=\left( \Omega _{e}-\Omega _{i}\right)
r/\alpha _{e}$ and $v_{ri}=\left( \Omega _{i}-\Omega _{e}\right) r/\alpha
_{i}$. Note that $\mathbf{\Delta v}_{e}=\mathbf{\Delta v}_{i}=\mathbf{0}$ is
exactly fulfilled. This allows to express the radial current $j$ = $%
nq(v_{ri}-v_{re})$ as
\begin{equation}
\mathbf{j}=nq\left( \Omega _{i}-\Omega _{e}\right) \left( \frac{1}{\alpha
_{i}}+\frac{1}{\alpha _{e}}\right) r\mathbf{e}_{r}\text{.}  \label{j}
\end{equation}
On the basis of these components, the electron and ion terms associated with
the centrifugal force, $\Delta v^{2}$ = $\Delta \left( v_{r}^{2}+v_{\theta
}^{2}\right) $ , can be expressed as 
\begin{equation}
\frac{\Delta {v_{e}}^{2}}{4}={\Omega _{e}}^{2}+\frac{\left( \Omega _{e}-\Omega
_{i}\right) ^{2}}{\alpha _{e}^{2}}\text{ , }\frac{\Delta {v_{i}}^{2}}{4}%
={\Omega _{i}}^{2}+\frac{\left( \Omega _{i}-\Omega _{e}\right) ^{2}}{\alpha
_{i}^{2}}\text{.}
\end{equation}
The final relations fulfilled by the vorticities $\Omega _{e/i}$ as a
function of the electric potential drive $\phi $ are thus given by 
\begin{subequations}
\begin{multline}
\left( \Omega _{e}-\Omega _{i}\right) \left( \frac{1}{\alpha _{e}}+\frac{1}{%
\alpha _{i}}\right) +\alpha _{e}\Omega _{e}+\frac{m_{e}}{m\nu }\left[ {\Omega
_{e}}^{2}+\frac{\left( \Omega _{e}-\Omega _{i}\right) ^{2}}{\alpha _{e}^{2}}%
\right] \\ = \frac{q}{2m\nu }\Delta \phi  \label{r1} 
\end{multline}
and
\begin{multline}
\left( \Omega _{i}-\Omega _{e}\right) \left( \frac{1}{\alpha _{e}}+\frac{1}{%
\alpha _{i}}\right) +\alpha _{i}\Omega _{i}+\frac{m_{i}}{m\nu }\left[ {\Omega
_{i}}^{2}+\frac{\left( \Omega _{i}-\Omega _{e}\right) ^{2}}{\alpha _{i}^{2}}%
\right] \\ = -\frac{q}{2m\nu }\Delta \phi \text{.}  \label{r2}
\end{multline}
\end{subequations}
$\Omega _{i/e}$ are the two unknowns of these two equations and the right
hand side of these relations, $\Delta \phi $, must be independent of the
radial position in order to find rigid body rotation solutions to this set
of algebraic equations Eqs.~(\ref{r1}) and (\ref{r2}). This can be achieved
if we consider a small uniform space charge $\rho =q\left(
n_{i}-n_{e}\right) $ ($n_{i}-n_{e}\ll n$). This uniform space charge is the
source of a linear electric field $E$ and a parabolic potential $\phi $ with
respect to $r$ illustrated in Fig.~\ref{Fig:Fig2}.

We define the $E$ cross $B$ angular velocity $\Omega =E/Br$ and the
frequency $\omega _{E}$ according to the definition 
\begin{equation}
\frac{{\omega _{E}}^{2}}{\omega _{i}\omega _{e}}=\frac{mE}{qB^{2}r}=\frac{%
m\Omega }{qB}\ll 1\text{,}
\end{equation}
such that Poisson's equation becomes : $\Delta \phi $ = $-2{\omega
_{E}}^{2}m/q $. With this definition of $\omega _{E}$ we have to solve a set of two coupled
fourth order algebraic equations, Eqs.~(\ref{alg1}) and (\ref{alg2}), with
two unknowns, the vorticities $\Omega _{i/e}$
\begin{subequations}
\begin{multline}
\left( \Omega _{e}-\Omega _{i}\right) \left( \alpha _{e}^{-1}+\alpha
_{i}^{-1}\right) +\alpha _{e}\Omega _{e}\\+\frac{m_{e}}{m\nu }\left[ {\Omega
_{e}}^{2}+\alpha _{e}^{-2}\left( \Omega _{e}-\Omega _{i}\right) ^{2}\right] +%
\frac{{\omega _{E}}^{2}}{\nu } = 0,  \label{alg1}
\end{multline}
\begin{multline}
\left( \Omega _{i}-\Omega _{e}\right) \left( \alpha _{e}^{-1}+\alpha
_{i}^{-1}\right) +\alpha _{i}\Omega _{i}\\+\frac{m_{i}}{m\nu }\left[ {\Omega
_{i}}^{2}+\alpha _{i}^{-2}\left( \Omega _{i}-\Omega _{e}\right) ^{2}\right] -%
\frac{{\omega _{E}}^{2}}{\nu } = 0.  \label{alg2}
\end{multline}
\end{subequations}
The solutions of this system of two algebraic equations describe the various
branches of collisional rigid body rotation in a fully ionized collisional
plasma. The solutions of Eqs.~(\ref{alg1}) and (\ref{alg2}), completed with relation
Eq.~(\ref{j}), provide the exact expression of the Ohmic current short
circuiting the G-J charge $\rho $.

Rather than an exact solution of the fourth order coupled equations Eqs.~(%
\ref{alg1}) and (\ref{alg2}), we seek an approximate expression of the
current based on a small collisionality expansion of the solution which
matches the Brillouin slow mode when $\nu \rightarrow 0$. The ordering
between the various time scales involved in these equations is : $\nu
_{i}\ll \nu \ll \nu _{e}\ll \Omega _{i}\sim \Omega _{e}\sim \Omega \ll
\omega _{i}\ll \omega _{e}$, thus we consider two small parameters : $m\nu
/qB$ = $\nu _{i}/\omega _{i}$ = $\nu _{e}/\omega _{e}$ $\ll 1$ and $m\Omega
/qB$ = $\Omega /\sqrt{\omega _{e}\omega _{i}}\ll 1$. We will solve Eqs.~(\ref
{alg1}) and (\ref{alg2}) through an expansion with respect to these two
small parameters.

The zero order solution is simply the $E$ cross \ $B$ flow without inertial
and collisional effects. The first order slow Brillouin solutions are the
sum of this azimuthal $E$ cross $B$ drift plus the first inertial drift
corrections 
\begin{subequations}
\begin{align}
\frac{\Omega _{e}}{\omega _{e}} & = -\frac{{\omega _{E}}^{2}}{\omega _{e}\sqrt{%
\omega _{i}\omega _{e}}}-\frac{{\omega _{E}}^{4}}{{\omega _{e}}^{3}\omega _{i}}%
+O\left( \frac{m\Omega }{qB}\right) ^{3}+O\left( \frac{m\nu }{qB}\right), \nonumber\\
  \label{br1} 
\end{align}
\begin{align}
\frac{\Omega _{i}}{\omega _{i}} & = -\frac{{\omega _{E}}^{2}}{\omega _{i}\sqrt{%
\omega _{i}\omega _{e}}}+\frac{{\omega _{E}}^{4}}{{\omega _{i}}^{3}\omega _{e}}%
+O\left( \frac{m\Omega }{qB}\right) ^{3}+O\left( \frac{m\nu }{qB}\right). \nonumber\\  \label{br2}
\end{align}
\end{subequations}
We can plug the expression of $\Omega _{i}-\Omega _{e}$ in Eqs.~(\ref{alg1})
and (\ref{alg2}) to get the next order and so on, but this level of
expansion is sufficient to analyze the nonlinear Ohmic current Eq.~(\ref{j})
as 
\begin{equation}
\frac{1}{\alpha _{e}}+\frac{1}{\alpha _{i}}=2\frac{m\nu }{qB}\left( \frac{%
\Omega _{e}}{\omega _{e}}+\frac{\Omega _{i}}{\omega _{i}}\right) \left[
1+O\left( \frac{m\Omega }{qB}\right) ^{2}\right] \text{.}  \label{br3}
\end{equation}

Using the relations Eqs.~(\ref{br1}), (\ref{br2}) and (\ref{br3}) and the
expression Eq.~(\ref{j}), \ the first order electric current $\mathbf{j}$,
with respect to an $m\Omega /qB\ll 1$ and $m\nu /qB\ll 1$ expansion, is
given by 
\begin{equation}
\frac{\mathbf{j}}{2nq}=\frac{\nu }{\sqrt{\omega _{i}\omega _{e}}}\frac{%
{\omega _{E}}^{4}}{{\omega _{i}}^{3}\omega _{e}}\Omega _{i}r\mathbf{e}_{r}=\frac{%
\nu _{i}\Omega ^{2}}{{\omega _{i}}^{3}}\frac{\mathbf{E}}{B}\text{.}
\label{jjj}
\end{equation}
As $\Omega \sim E$, this Ohmic current $j$ is a cubic function of the
electric field $E$, its origin, analyzed in Sec.~\ref{Sec:Sec4}, is to be traced back
to an interplay between Coriolis, centrifugal and Coulomb frictions forces.
The current Eq.~(\ref{jjj}) scales as $B^{-6}$ with respect to the magnetic
field strength, this scaling explains the high efficiency of rotation
generation in confined plasmas identified and analyzed within the context of
wave driven magneto-electric confinement studies~\cite{Rax2017}.

To summarize this section, the weak dissipative current $\mathbf{j}$,
expressed by Eq.~(\ref{jjj}), continuously short circuits the G-J charges $%
\rho $ between field lines. To evaluate this space charge $\rho $, and
validate the ordering $\rho \ll nq$, we apply Gauss theorem to this
cylindrical configuration, $\rho /\varepsilon _{0}\varepsilon _{\bot }$ = $%
2E/r$, to get the expression 
\begin{equation}
\frac{n_{i}-n_{e}}{n_{i}+n_{e}}=\frac{\rho }{2nq}=\frac{\varepsilon
_{0}\varepsilon _{\bot }\Omega B}{nq}=\frac{{\Omega }}{{\omega }_{i}}\ll 1%
\text{,}
\end{equation}
which is coherent with the fact that we consider the slow branch (${\Omega }%
\ll {\omega }_{i}$) of the two fast and slow Brillouin rotation modes~\cite{Brillouin1945,Davidson2001}.

Note the Maxwell time $\tau _{M}$ associated with the resistive decay of the
uncompensated free charges $\rho $ is no longer given by the classical
expression $\tau _{M}$ $\sim $ $\left| \rho /\mathbf{\nabla }\cdot \mathbf{j}%
\right| $ $\sim $ $\nu _{e}/\omega _{p}^{2}$ but is given by the ratio: $%
\tau _{M}$ $\sim $ $\left| \rho /\mathbf{\nabla }\cdot \mathbf{j}\right| $ $%
\sim $ $\omega _{i}^{2}/\Omega ^{2}\nu _{i}$. Thus, (\textit{i}) along the
field lines the relaxation time of free charges scales as $\nu _{e}$, and (%
\textit{ii}) across the field lines as $\left( \omega _{i}/\Omega \right)
^{2}/\nu _{i}$, this is also the classical behavior, $\nu _{e}$ versus $%
1/\nu _{i}$, of both mobility and diffusion (\textit{i}) along and (\textit{%
ii}) across the magnetic field. A complete analysis of the dynamics of the
G-J charges relaxation will be presented in Secs.~\ref{Sec:Sec5} and \ref{Sec:Sec6} where the \textit{%
quality factor} of angular momentum storage and the \textit{efficiency} of
angular momentum generation will give two practical characterizations of
this relaxation besides the Maxwell time.

Although the occurrence of dissipative inertial effects of the type Eq.~(\ref
{ohoh2}) in accelerated flows is described in the literature~\cite{Rozhansky2008,Landau1984,Helander2002}, the
analysis and expansion of the inertial term $d\mathbf{v}/dt$ = $\left( 
\mathbf{v}\cdot \mathbf{\nabla }\right) \mathbf{v}$ with a rotating two
fluid model leading to the general nonlinear conductivity Eq.~(\ref{jjj})
did not appear in the literature. The relaxation of a fully ionized
collisional Brillouin flow did not attract much attention and the exact
solution of the weakly ionized collisional Brillouin flow appears only
recently in the literature~\cite{Rax2015}. Eq.~(\ref{jjj}) is therefore one of the new
results of this study. Finally, while Eq.~(\ref{jjj}) has been derived assuming rigid body rotation and cylindrical geometry, we will expose in Sec.~\ref{Sec:Sec4} that this result can be recovered and generalized in the more general case of an axisymmetric rotating flow fullfiling Ferraro isorotation. As it will be shown, the only difference between Eq.~(\ref{jjj}) and the generic case is a geometrical factor associated with the radial profile of the magnetic flux function.



\section{Isorotation of an axisymmetric plasma}
\label{Sec:Sec3}

\begin{figure*}
\begin{center}
\includegraphics[width = 14cm]{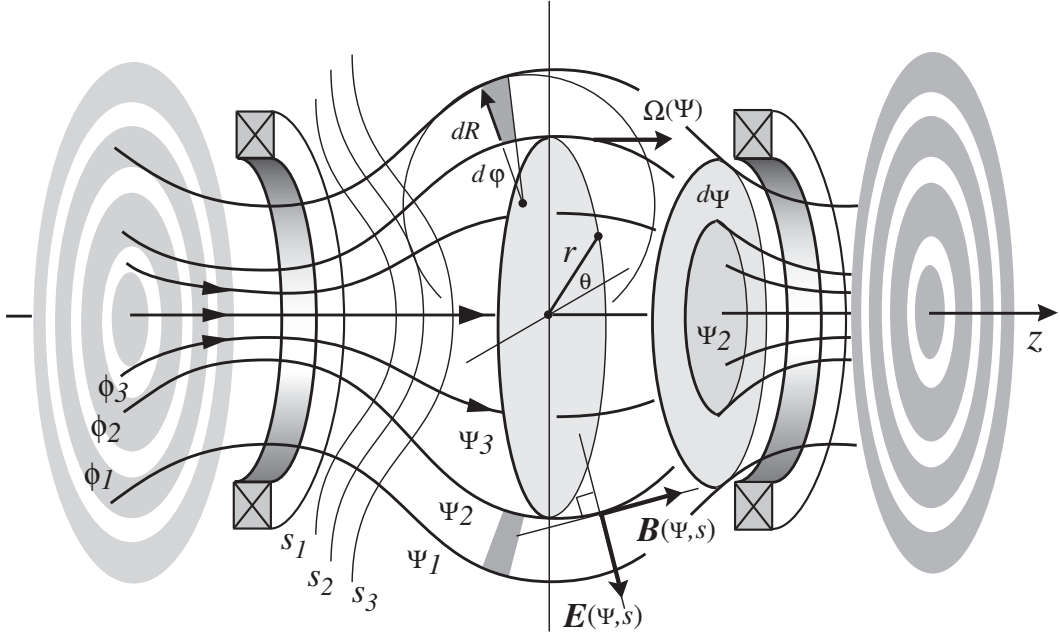}
\caption{Rotating axisymmetric magnetized plasma, $R$ is the radius of
curvature, $\phi $ is the electric potential, $\Psi $ the magnetic flux and $%
\Omega \left( \Psi \right) $ the angular velocity, $\left( \Psi ,\theta
,s\right) $ or $\left( \phi ,\theta ,s\right) $ provides a set of
generalized coordinates. }
\label{Fig:Fig3}
\end{center}
\end{figure*}

In order to set the frame for a generalization of the expression of the
nonlinear Ohmic current Eq.~(\ref{jjj}), we consider a fully ionized, steady
state, axisymmetric, plasma depicted in Fig.~\ref{Fig:Fig3}. We use cylindrical polar
coordinates $\left( r,\theta ,z\right) $ associated with the cylindrical
polar basis $\left[ \mathbf{e}_{r},\mathbf{e}_{\theta },\mathbf{e}%
_{z}\right] $. The structure of this general axisymmetric configuration
around the $z$ axis is described by the electric field $\mathbf{E}$ and
magnetic field $\mathbf{B}$%
\begin{equation}
\mathbf{B}\left( r,z\right) =B_{r}\mathbf{e}_{r}+B_{z}\mathbf{e}_{z}\text{ , 
}\mathbf{E}\left( r,z\right) =E_{r}\mathbf{e}_{r}+E_{z}\mathbf{e}_{z}\text{.}
\end{equation}
We introduce the magnetic flux function $2\pi \Psi \left( r,z\right) $ and
the electrostatic potential $\phi \left( r,z\right) $ to describe these
magnetic and electric fields. With these flux and potential functions the
expressions of these fields become
\begin{equation}
\mathbf{B}\left( r,z\right) =\mathbf{\nabla }\Psi \times \frac{\mathbf{e}%
_{\theta }}{r}\text{ , }\mathbf{E}\left( r,z\right) =-\mathbf{\nabla }\phi 
\text{.}
\end{equation}
Because of the very large conductivity along the field lines, $\mathbf{E\cdot
B}=0$, and the electric field is everywhere perpendicular to the magnetic
field. Magnetic surfaces $\Psi \left( r,z\right) =C^{te}$ are therefore
equipotential surfaces $\phi \left( r,z\right) =C^{te}$. As a result $\phi $ is
determined by the magnetic flux $\Psi $: $\phi \left( r,z\right) =\phi
\left( \Psi \right) $. When we will consider diamagnetic flow we will also
assume that the pressure gradient is perpendicular to the magnetic surfaces, \emph{i.~e. } $\Psi \left( r,z\right) $ are isobaric and isopotential surfaces. 

Both electrons and ions rotate around the $z$ axis with an $E$ cross $B$ velocity 
$\mathbf{v}$ defining the angular velocity $\Omega \left( r,z\right) $, 
\begin{equation}
\mathbf{v}\left( r,z\right) =\frac{\mathbf{E}\times \mathbf{B}}{B^{2}}%
=\Omega \left( r,z\right) r\mathbf{e}_{\theta }\text{.}
\end{equation}
We introduce $s$ the curvilinear abscissa along each field line such that $%
\left( \Psi ,\theta ,s\right) $ provide a set of coordinates illustrated on
Fig.~\ref{Fig:Fig3}. Let us call $dR$ the small perpendicular distance between two neighbouring
magnetic surfaces. According to Fig.~\ref{Fig:Fig3}, the conservation of magnetic flux
can be written $B2\pi rdR=2\pi d\Psi $, and the relation between electric
field and potential is $EdR=-d\phi $. The elimination of $dR$ between these
two relations leads to the \textit{Ferraro isorotation law}
\begin{equation}
\Omega \left( \Psi \right) =\frac{d\phi }{d\Psi }\text{.}  \label{fere}
\end{equation}
Each magnetic surface can be viewed as an equipotential surface rotating
uniformly and the full configuration as a set of nested magnetic surfaces
rotating around the $z$ axis~\cite{Ferraro1937}.

We assume that the temperature difference $\Delta T=T_{1}-T_{2}$ between two
magnetic surfaces $\Psi _{1}$ and $\Psi _{2}$ is smaller than the voltage
drop $\Delta \phi =\phi _{1}-\phi _{2}$ $\ $between these two surfaces
according to the ordering : $k_{B}\Delta T\ll q\Delta \phi $. Thus the
pressure force can be neglected in front of the electric force as $\left| 
\mathbf{\nabla }nk_{B}T\right| \ll nq\left| \mathbf{\nabla }\phi \right| $.
This strong ordering is fulfilled by the fast rotating discharges of
interest for thermonuclear fusion and allows to study separately inertial
effects and finite Larmor radius effects. For the weak ordering, $\left| 
\mathbf{\nabla }nk_{B}T\right| \sim nq\left| \mathbf{\nabla }\phi \right| $,
the pressure profiles $P\left( \Psi \right) $ and $\phi \left( \Psi \right) $
provide a set of given data to solve a Grad-Shafranov equation for $\Psi
\left( r,z\right) $ similar to the profiles $P\left( \Psi \right) $ and $%
I\left( \Psi \right) $ for the classical Grad-Shafranov equation. This
analysis of the mechanical/electrical MHD equilibrium is far beyond the
scope of this study which is only devoted to the analysis of Ohmic
dissipation for a given configuration $\Psi \left( r,z\right) $ and $\phi
\left( \Psi \right) $.

Before analyzing the collisional nonlinear short circuiting of the G-J
charges $\rho $, we will demonstrate that they are completely determined by
the magnetic flux $\Psi \left( r,z\right) $ and the electric potential $\phi
\left( \Psi \right) $. In the historical papers on pulsar magnetosphere the
G-J charges were usually derived starting from Amp\`{e}re and Poisson
equations associated with $\Psi $ and $\phi $~\cite{Goldreich1969,Julian1973}. Here we will
introduce the radius of curvature of the field lines, $R^{-1}=\left| d%
\mathbf{b}/ds\right| $, to get new simple expressions. Let us call $R\left(
\Psi ,s\right) $ the radius of curvature of a field line and $\varphi $ the
angle along the osculating circle. Consider, in Fig.~\ref{Fig:Fig3}, a small grey box $%
\left( dR,Rd\varphi ,2\pi r\right) $ between two magnetic surfaces: $dR$ is
along $\mathbf{E}$, $Rd\varphi $ along $\mathbf{B}$ and $2\pi r$ along $%
\mathbf{e}_{\theta }$. We use Gauss theorem, $\left( E+dE\right) \left(
R+dR\right) d\varphi $ $-$ $ERd\varphi =\rho Rd\varphi dR/\varepsilon
_{0}\varepsilon _{\bot }$ and the relation between electric field and
potential, $EdR=-d\phi $, to express the G-J charge density $\rho $ as
\begin{equation}
\frac{\rho }{\varepsilon _{0}}=-\frac{{c}^{2}}{{V_{A}}^{2}}\frac{E}{\Omega R}%
\frac{\partial RE}{\partial \Psi }\text{,}  \label{gjgl}
\end{equation}
where $V_{A}$ is Alfven's velocity and $c$ the velocity of light. Consider,
in Fig.~\ref{Fig:Fig3}, a small closed loop $\left( \pm dR,\pm Rd\varphi \right) $ along
two field lines and between two magnetic surfaces, $dR$ is along $\mathbf{E}$
and $Rd\varphi $ along $\mathbf{B}$. Amp\`{e}re's theorem can be written $%
BRd\varphi $ $-$ $\left( B+dB\right) $ $\left( R+dR\right) d\varphi $ $=-\mu
_{0}Rd\varphi dR\rho E/B$ where we have neglected the diamagnetic currents
in front of the $E$ cross $B$ convection current. The G-J charges density $%
\rho $ is thus given \ as a function of the magnetic field as
\begin{equation}
\frac{\rho }{\varepsilon _{0}}=-c^{2}\frac{B}{\Omega R}\frac{\partial RB}{%
\partial \Psi }\text{.}
\end{equation}
Note that $R=\infty $ for the Brillouin flow. 

To summarize this brief
presentation of a generic axisymmetric rotating plasma: the magnetic field
is structured by magnetic surfaces $\Psi \left( r,z\right) $ such that $%
\mathbf{B}\left( \Psi ,s\right) =\mathbf{\nabla }\Psi \times \mathbf{e}%
_{\theta }/r$, the electric field $\mathbf{E}\left( \Psi ,s\right) =-\Omega
\left( \Psi \right) \mathbf{\nabla }\Psi $ is everywhere perpendicular to
the magnetic field and the plasma flow fulfils the Ferraro isorotation law $%
\mathbf{v}\left( \Psi ,s\right) =\Omega \left( \Psi \right) r\mathbf{e}%
_{\theta }$. For this classical flow, as electrons and ions velocities are
equal, there is no Coulomb friction between electrons and ions and thus no
Ohmic collisional dissipation of this equilibrium described by two given
functions $\left[ \Psi \left( r,z\right) \text{, }\phi \left( \Psi \right)
\right] $ or equivalently by $\left[ \Psi \left( r,z\right) \text{, }\Omega
\left( \Psi \right) \right] $.

\section{Nonlinear Ohmic conduction in a Ferraro flow}
\label{Sec:Sec4}

In this section we consider a two fluids model but we adopt a Lagrangian
point of view, different from the Eulerian point of view used in Sec.~\ref{Sec:Sec2}.
Moreover, taking advantage of \ Ferraro isorotation law Eq.~(\ref{fere}), we
consider the dynamic on a magnetic surface $\Psi $ in the frame corotating
at the angular velocity $\Omega \left( \Psi \right) $, different from the
laboratory frame point of view used in Sec.~\ref{Sec:Sec2}.

In this (co)rotating frame the electron and ion velocities are $\mathbf{v}%
_{e}$ and $\mathbf{v}_{i}$, the electric field cancels everywhere on the $%
\Psi $ magnetic surface, but we have to take into account the Coriolis and
centrifugal forces to write down electrons and ions forces balances
\begin{subequations}
\begin{multline}
m_{e}\Omega ^{2}r\mathbf{e}_{r}+2m_{e}\mathbf{v}_{e}\times \Omega \mathbf{e}%
_{z}-q\mathbf{v}_{e}\times B\mathbf{b}  \\ =m_{e}\nu _{e}\left( \mathbf{v}_{e}-%
\mathbf{v}_{i}\right), \label{fff1} 
\end{multline}
\begin{multline}
m_{i}\Omega ^{2}r\mathbf{e}_{r}+2m_{i}\mathbf{v}_{i}\times \Omega \mathbf{e}%
_{z}+q\mathbf{v}_{i}\times B\mathbf{b} \\ =m_{i}\nu _{i}\left( \mathbf{v}_{i}-%
\mathbf{v}_{e}\right), \label{fff2}
\end{multline}
\end{subequations}
where the local unit vector $\mathbf{b}=\mathbf{B}/B$ is given by 
\begin{equation}
Br\mathbf{b}\left( r,z\right) =\frac{\partial \Psi }{\partial r}\mathbf{e}%
_{z}-\frac{\partial \Psi }{\partial z}\mathbf{e}_{r}\text{.}
\end{equation}
The first terms on the left hand sides of Eqs.~(\ref{fff1}) and (\ref{fff2})
are the centrifugal and Coriolis forces and the right hand sides describe
collisional friction. Introducing the cyclotron frequencies we have to solve
the following set of equations for the velocities $\mathbf{v}_{e}$ and $%
\mathbf{v}_{i}$ on the flux surface $\Psi $ in the frame rotating at angular
velocity $\Omega \left( \Psi \right) $%
\begin{subequations}
\begin{align}
2\mathbf{v}_{e}\times \frac{\Omega }{\nu _{e}}\mathbf{e}_{z}-\mathbf{v}%
_{e}\times \frac{\omega _{e}}{\nu _{e}}\mathbf{b}-\mathbf{v}_{e}+\mathbf{v}%
_{i} &=-\frac{\Omega ^{2}}{\nu _{e}}r\mathbf{e}_{r}\text{,}  \label{ion}\\
2\mathbf{v}_{i}\times \frac{\Omega }{\nu _{i}}\mathbf{e}_{z}+\mathbf{v}%
_{i}\times \frac{\omega _{i}}{\nu _{i}}\mathbf{b}-\mathbf{v}_{i}+\mathbf{v}%
_{e} &=-\frac{\Omega ^{2}}{\nu _{i}}r\mathbf{e}_{r}\text{,}
\end{align}
\end{subequations}
under the strong ordering: $\nu _{i}\ll \nu $ $\ll \nu _{e}\ll \Omega $ $%
\ll \omega _{i}\ll \omega _{e}$, $v_{e}\sim v_{i}\ll \Omega r$. Note that
in the lab frame $\nu _{i}\ll \nu $ $\ll \nu _{e}\ll \Omega $ $\ll \omega
_{i}$ $\ll \omega _{e}$ but $v_{e}$ $\sim $ $v_{i}\sim \Omega r$. Following
the ordering, the ion dynamics Eq.~(\ref{ion}) is dominated by the balance
between the magnetic Laplace force and the inertial centrifugal force : $%
\Omega ^{2}r\mathbf{e}_{r}$ $=$ $-\omega _{i}\mathbf{v}_{i}\times \mathbf{b}$%
. This last equation is solved as a guiding center centrifugal force drift 
\begin{equation}
\mathbf{v}_{i}=-\frac{\Omega ^{2}}{\omega _{i}}r\mathbf{b}\times \mathbf{e}%
_{r}\text{.}  \label{firion}
\end{equation}

Eq.~(\ref{firion}) describes an ion collisionless azimuthal cross-field flow
driven by the centrifugal force. The same azimuthal collisionless flow \ for
the electrons is smaller by an $m_{e}/m_{i}$ mass ratio factor. Then we have
to consider the Coulomb collisions and the friction force of the ions with
azimuthal velocities $\mathbf{v}_{i}$ with the electrons with azimuthal
velocities $\mathbf{v}_{e}=m_{e}\mathbf{v}_{i}/m_{i}$. This collisional
coupling is the source of an azimuthal force $\mathbf{F}_{i}$ on the ions
and $\mathbf{F}_{e}\left( =-\mathbf{F}_{i}\right) $ on the electrons 
\begin{equation}
\mathbf{F}_{i}=-m_{i}\nu _{i}\mathbf{v}_{i}=m_{i}\nu _{i}\frac{\Omega ^{2}}{%
\omega _{i}}r\mathbf{b}\times \mathbf{e}_{r}\text{.}
\end{equation}
These azimuthal friction forces are the source of a cross-field ion flow $%
\mathbf{V}_{i}$ and a cross-field electron flow $\mathbf{V}_{e}\left( =%
\mathbf{V}_{i}\right) $, thus no net electric current is observed at this
level of analysis 
\begin{equation}
\mathbf{V}_{i}=\frac{\mathbf{F}_{i}\times \mathbf{B}}{qB^{2}}=\frac{\nu
_{i}\Omega ^{2}}{\omega _{i}^{2}}r\left( \mathbf{b}\times \mathbf{e}%
_{r}\right) \times \mathbf{b}\text{.} \label{ambi}
\end{equation}

Now we take into account the Coriolis force, which has been neglected up to
now according to the ordering. The two small collisional flows $\mathbf{V}%
_{e}$ and $\mathbf{V}_{i}$ are responsible for small azimuthal Coriolis
forces $\mathbf{f}_{i}$ and $\mathbf{f}_{e}\left( \ll \mathbf{f}_{i}\right) $
\begin{equation}
\mathbf{f}_{i}=2m_{i}\mathbf{V}_{i}\times \Omega \mathbf{e}_{z}=2m_{i}\Omega
\nu _{i}\frac{\Omega ^{2}}{{\omega _{i}}^{2}}r\left[ \left( \mathbf{b}\times 
\mathbf{e}_{r}\right) \times \mathbf{b}\right] \times \mathbf{e}_{z}\text{.}
\label{cor}
\end{equation}
This Coriolis/collisional force drives a cross-field ion flow $\mathbf{w}%
_{i} $ and a cross-field electron flow $\mathbf{w}_{e}\left( \ll \mathbf{w}%
_{i}\right) $%
\begin{equation}
\mathbf{w}_{i}=\frac{\mathbf{f}_{i}\times \mathbf{B}}{qB^{2}}=2\nu _{i}\frac{%
\Omega ^{3}}{{\omega _{i}}^{3}}r\left( \left[ \left( \mathbf{b}\times \mathbf{e%
}_{r}\right) \times \mathbf{b}\right] \times \mathbf{e}_{z}\right) \times 
\mathbf{b}\text{.}  \label{dert}
\end{equation}
The direction of this flow is along the electric field and it is
proportional to the momentum exchange frequency 
\begin{align}
\left( \left[ \left( \mathbf{b}\times \mathbf{e}_{r}\right) \times \mathbf{b}%
\right] \times \mathbf{e}_{z}\right) \times \mathbf{b} & =\left( \frac{1}{rB}%
\frac{\partial \Psi }{\partial r}\frac{\mathbf{E}}{E}\times \mathbf{e}%
_{z}\right) \times \mathbf{b}\nonumber\\ 
 & =\frac{1}{r^{2}B^{2}}\left( \frac{\partial \Psi 
}{\partial r}\right) ^{2}\frac{\mathbf{E}}{E}\text{.}
\end{align}
The associated electric current $\mathbf{j}=nq\mathbf{w}_{i}$ is the
nonlinear Ohmic current providing G-J charge relaxation. Its final
expression is given by 
\begin{equation}
\frac{\mathbf{j}}{2nq}=\frac{\nu _{i}}{E^{2}B}\frac{\Omega ^{4}}{{\omega
_{i}}^{3}}\left( \frac{\partial \Psi }{\partial r}\right) ^{2}\mathbf{E}=%
\frac{\nu _{i}\Omega ^{2}}{{\omega _{i}}^{3}}\frac{E_{r}^{2}}{E^{2}}\frac{%
\mathbf{E}}{B}\text{,}  \label{curent}
\end{equation}
where we have used the Ferraro isorotation law Eq.~(\ref{fere}) to express
the right hand side of this relation.

Moving back to the laboratory frame does not change this expression of the
radial Ohmic current Eq.~(\ref{curent}) which is one of the main new results
of this study. For a Brillouin flow $E_{r}=E$ we recover exactly the result
Eq.~(\ref{jjj}) obtained at the end of Sec.~\ref{Sec:Sec2} through a careful expansion
of the solutions of the exact fourth order coupled equations fulfilled by
the vorticities in the laboratory frame. As $\Omega \sim E$ the conductivity
scales as $E^{2}$ and is clearly nonlinear.

To compare this new result $\mathbf{j}$ $=$ $\sigma _{NL}\mathbf{E}$
described by Eq.~(\ref{curent}) with the usual picture on Ohmic dissipation we
introduce the classical Braginsky expression for the electric conductivity
perpendicular to the magnetic field~\cite{Braginskii1965}, $\sigma _{B}=nq^{2}\nu
_{e}/m_{e}{\omega _{e}}^{2}$, to get the scaling and ordering 
\begin{equation}
\frac{\sigma _{NL}}{\sigma _{B}}=2\left( \frac{\Omega }{\omega _{i}B}\right)
^{2}\left( \frac{1}{r}\frac{\partial \Psi }{\partial r}\right) ^{2}\sim
\left( \frac{\Omega }{\omega _{i}}\right) ^{2}\ll 1\text{.}
\end{equation}

To summarize Secs.~\ref{Sec:Sec3} and \ref{Sec:Sec4}, we have demonstrated that a small breakdown
of quasineutrality described by the G-J charges Eq.~(\ref{gjgl}) leads to a
(iso)rotation of the axisymmetric plasma and that these charges are short
circuited by a nonlinear Ohmic current Eq.~(\ref{curent}) resulting from an
interplay between (\textit{i}) Coriolis, (\textit{ii}) centrifugal and (%
\textit{iii}) Coulomb friction forces. To complete this description of
charges we have also identified through Eq.~(\ref{ambi}) an outward ambipolar particle flow $n\mathbf{V}_{i}+n\mathbf{V}_{e}$, perpendicular to
the magnetic surfaces such that the associated mass flow $\mathbf{J}$ is
given by
\begin{equation}
\frac{\mathbf{J}}{nm_{i}}=\nu _{i}\frac{\Omega ^{2}}{{\omega _{i}}^{2}}r\left( 
\mathbf{b}\times \mathbf{e}_{r}\right) \times \mathbf{b}=s_{\pm }\frac{\nu
_{i}\Omega }{{\omega _{i}}^{2}}\frac{B_{z}}{B}\frac{\mathbf{E}}{B}\text{,}
\label{ambi2}
\end{equation}
where $s_{\pm }$ is the sign of $-\left( \mathbf{\Omega }\cdot \mathbf{e}%
_{z}\right) \left( \mathbf{E}\cdot \mathbf{e}_{r}\right) \left( \mathbf{B}%
\cdot \mathbf{e}_{z}\right) $. Note that for the Brillouin flow the
expression $\left( \mathbf{b}\times \mathbf{e}_{r}\right) \times \mathbf{b}$
= $\mathbf{e}_{r}$ clearly shows that this ambipolar mass flux is radially outward. Moving back to the laboratory frame does not change the
expression of this collisional radial ambipolar flow Eq.~(\ref{ambi2}).

\section{Quality factor of angular momentum storage}
\label{Sec:Sec5}

As depicted in Fig.~\ref{Fig:Fig3}, axisymmetric plasma rotation can be sustained
through the polarization of each magnetic surface $\Psi $ with a system of
concentric conductive electrodes intercepting the magnetic field lines at
the left and right edges of the plasma. A voltage generator is used to
sustain a voltage drop $\Delta \phi $ in between the magnetic surfaces $\Psi 
$ and $\Psi +\Delta \Psi $ resulting in a plasma rotation $\Omega =\Delta
\phi /\Delta \Psi $ of these surfaces according to the Ferraro isorotation law
Eq.~(\ref{fere}).

This voltage drop $\Delta \phi $ is short circuited by the inertial Ohmic
current Eq.~(\ref{curent}) and a steady state active power consumption takes
place to sustain the rotation. The reactive power is associated with energy
storage during the initial transient build up of the polarization and
rotation. The rotating plasma discharge can be viewed either as an
electrostatic energy storage, or as a kinetic energy storage, with total
energy $U_{\Omega }$ \ given by 
\begin{align}
U_{\Omega } & =\iiint \frac{\rho \phi }{2}d\tau =\iiint \varepsilon _{\bot
}\varepsilon _{0}\frac{E^{2}}{2}d\tau \nonumber\\ & =\iiint nm_{i}\frac{\Omega ^{2}r^{2}}{2%
}d\tau \text{,}  \label{UU}
\end{align}
where $d\tau $ is the volume element and the integrations run over the
volume of the plasma. Note that $\Omega r=E/B$ and $\varepsilon _{\bot }$ = $%
{\omega _{pi}}^{2}/{\omega_{i}}^{2}$ leads directly the last identity of
Eq.~(\ref{UU}).

This global energy content $U_{\Omega }$ \ is dissipated through
electron-ion collisions. If the power and control systems are switched off,
rotation slows down and the density decays because : (\textit{i}) on each
magnetic surface the angular momentum around $z$ is continuously destroyed
by the resistive Ohmic torque associated with the force $\mathbf{j}\times 
\mathbf{B}$, Eq. (\ref{curent}), and (\textit{ii}) energy leaks out of the
magnetic surface through the outward convection of energy\ associated with
the ambipolar mass flow $\mathbf{J}$, Eq.~(\ref{ambi2}).

At a given point $r$ the decay of the density of kinetic energy is the sum
of two terms : (\textit{i}) the resistive work of the density of force $%
\mathbf{j}$ $\times \mathbf{B}$, and (\textit{ii}) the convective ambipolar
mass flux described by $\mathbf{J}$%
\begin{equation}
\frac{\partial }{\partial t}\left( n\frac{m_{i}}{2}\Omega ^{2}r^{2}\right)
=nm_{i}\Omega \frac{\partial \Omega }{\partial t}r^{2}+\frac{\partial n}{%
\partial t}\frac{m_{i}}{2}\Omega ^{2}r^{2}\text{.}  \label{ener}
\end{equation}
The decay of the angular velocity $\partial \Omega /\partial t$ is due to
the dissipative work of the torque $\mathbf{\mathbf{r}\times }\left( \mathbf{%
j}\times \mathbf{B}\right) $ and the decay of the density $m_{i}\partial
n/\partial t$ $\ $= $-\mathbf{\nabla }\cdot \mathbf{J}$ is due to the
ambipolar mass flux $\mathbf{J}$%
\begin{subequations}
\begin{align}
nm_{i}\Omega \frac{\partial \Omega }{\partial t}r^{2} =&-\mathbf{\mathbf{r}%
\times }\left( \mathbf{j}\times \mathbf{B}\right) \cdot \mathbf{\Omega }%
\text{,} \\
\frac{m_{i}}{2}\frac{\partial n}{\partial t}\Omega ^{2}r^{2} =&-\frac{%
\Omega ^{2}r^{2}}{2}\mathbf{\nabla }\cdot \mathbf{J}\text{.}
\end{align}
\end{subequations}
The rate of collisional energy dissipation $dU_{\Omega }/dt$ is thus given
as a function of $\ \mathbf{j}$ and $\mathbf{J}$, Eqs.~(\ref
{curent}) and (\ref{ambi2}), by 
\begin{equation}
\frac{dU_{\Omega }}{dt}=-\iiint \left[ \mathbf{\mathbf{r}\times }\left( 
\mathbf{j}\times \mathbf{B}\right) \cdot \mathbf{\Omega }+\frac{\Omega
^{2}r^{2}}{2}\mathbf{\nabla }\cdot \mathbf{J}\right] d\tau \text{,}
\label{pow}
\end{equation}
where integration runs over the volume of the plasma. In this section,
to establish the expression of the quality factor, we will use the Brillouin
flow approximation $\partial \Psi /\partial r=rB_{z}\approx rB$ or $\partial
\Psi /\partial r=E_{r}/\Omega \approx E/\Omega $ so that the expressions of
the collisional electric current flow Eq. (\ref{curent}) and the collisional
ambipolar particle flow Eq. (\ref{ambi2}) become 
\begin{subequations}
\begin{align}
\mathbf{j}&=2nq\frac{\nu _{i}\Omega ^{2}}{{\omega _{i}}^{3}}\frac{\mathbf{E}}{B}%
\text{,}\\
\mathbf{J}&=nm_{i}\frac{\nu _{i}\Omega }{{\omega _{i}}^{2}}\frac{E}{B}%
\mathbf{e}_{r}\text{.}
\end{align}
\end{subequations}
Note that $\mathbf{\nabla }\cdot \mathbf{J}=2nm_{i}\nu _{i}\Omega
^{2}/{\omega _{i}}^{2}$ and $\mathbf{\nabla }\cdot \mathbf{j}$ $=$ $4nq\nu
_{i}\Omega ^{3}/{\omega _{i}}^{3}$ are independent of the radius $r$. The
generalization of the results obtained in this section to a generic Ferraro
flow requires the inclusion of $\partial \Psi /\partial r$ and $%
\partial \Psi /\partial z$ \ factors associated with the full expressions
given by Eqs. (\ref{curent}) and (\ref{ambi2}). \ With these simple
expressions, the power of the resistive torque and the energy convection by
the ambipolar mass flow are given by 
\begin{subequations}
\begin{align}
\mathbf{\mathbf{r}\times }\left( \mathbf{j}\times \mathbf{B}\right) \cdot 
\mathbf{\Omega } &= 2qn\nu _{i}\frac{\Omega ^{4}B}{{\omega _{i}}^{3}}r^{2}%
\text{,}  \label{torq} \\
\frac{\Omega ^{2}r^{2}}{2}\mathbf{\nabla }\cdot \mathbf{J} &= qn\nu _{i}%
\frac{\Omega ^{4}B}{{\omega _{i}}^{3}}r^{2}\text{.}  \label{flowr}
\end{align}
\end{subequations}
Thus one third of the energy decay is due to the collisional particles
losses and two thirds to the nonlinear Ohmic conductivity. The ratio of the
energy decay during one turn $-\left( dU_{\Omega }/dt\right) \left( 2\pi
/\Omega \right) $ to the stored energy $U_{\Omega }$ defines the \textit{%
quality factor }$Q$ of this energy storage system 
\begin{equation}
\frac{1}{Q}=-\frac{2\pi dU_{\Omega }/dt}{\Omega U_{\Omega }}=12\pi \frac{%
\iiint q\nu _{i}\Omega ^{4}B\omega _{i}^{-3}r^{2}d\tau }{\Omega \iiint
m_{i}\Omega ^{2}r^{2}d\tau }=12\pi \frac{\nu _{i}\Omega }{{\omega _{i}}^{2}}%
\text{.}  \label{qqq}
\end{equation}
The time scale for collisional relaxation is not $1/\nu _{i}$ but far longer
by a factor $\left( \omega _{i}/\Omega \right) ^{2}$ already identified for
the evaluation of the Maxwell time at the end of Sec.~\ref{Sec:Sec2}. 

Rather than using the resistive torque $\mathbf{\mathbf{r}\times }\left( 
\mathbf{j}\times \mathbf{B}\right)$ Eq.~(\ref{torq}) to evaluate the power
of the Ohmic current Eq.~(\ref{curent}) short circuiting the G-J charges, we
can consider the rate of resistive charge depletion $\partial \rho /\partial
t$ given by $\partial \rho /\partial t=-\mathbf{\nabla }\cdot \mathbf{j}$.
This Ohmic charge decay is then used to express the power
\begin{equation}
-\iiint \phi \frac{\partial \rho }{\partial t}d\tau =\iiint 2nq\nu _{i}\frac{%
\Omega ^{4}}{{\omega _{i}}^{3}}Br^{2}d\tau \text{,}
\end{equation}
where the electrostatic potential is $\phi =-Er/2=-B\Omega r^{2}/2$. We
recover Eq.~(\ref{torq}) which confirms that, as expected, the electrical and mechanical power
balances give the very same result.

\section{Efficiency of RF angular momentum generation}
\label{Sec:Sec6}

The previous section was devoted to the study of the free decay of a
rotating plasma column after all the sustaining and control systems have been
switched off. 

In this section we consider instead a situation where the plasma column is
sustained in steady state rotation (\textit{i}) by an RF wave system,
providing angular momentum injection, and (\textit{ii}) by a pellet, or a
gas puffing, system providing particle fuelling to maintain a steady state
density profile. We thus assume that there is a source of particles on each magnetic surface 
$S\left( \Psi \right) $ such that $\partial n/\partial t=0$. When an
ion at rest appears on the surface $\Psi $ from neutral ionization, it
starts to rotate at the velocity $r\Omega \left( \Psi \right) =E/B$, and
compensates the collisional ambipolar depletion associated with $\mathbf{J}$
given in Eq. (\ref{ambi2}).

Instead of polarizing end plates electrodes with a
voltage generator, the angular momentum of the axisymmetric discharge
depicted in Fig.~\ref{Fig:Fig3} can be sustained using RF waves carrying angular
momentum. To ensure continuous angular momentum input, resonant and dispersive conditions for these waves to propagate and be absorbed within the plasma can be identified.


To simplify the analysis, we limit ourselves here to the Brillouin flow
approximation $\partial \Psi /\partial r=rB_{z}\approx rB$ or $\partial \Psi
/\partial r=E_{r}/\Omega \approx E/\Omega $. Note though that the generalization of the
results to a generic Ferraro flow simply requires including the $%
\partial \Psi /\partial r$ and $\partial \Psi /\partial z$ factors. The
resistive torque $\mathbf{r}\times\left( \mathbf{j}\times \mathbf{B}%
\right) $ slows down rotation while the absorption of
angular momentum from the wave spins up rotation. If we call $dL/dt$
the rate of wave angular momentum absorption per particle, the angular
momentum balance between absorption and dissipation is given by 
\begin{equation}
\frac{dL}{dt}=\mathbf{\mathbf{r}\times }\frac{\mathbf{j}\times \mathbf{B}}{n}%
\cdot \mathbf{e}_{z}=2q\nu _{i}\frac{\Omega ^{3}B}{{\omega _{i}}^{3}}r^{2}%
\text{.}  \label{J1}
\end{equation}

When a wave is absorbed by a plasma, (\textit{i}) energy, (\textit{ii})
linear momentum and (\textit{iii}) angular momentum are transferred from the
wave to the particles. Consider (\textit{i}) a cylindrical wave, $f\left(
r\right) \sin (l\theta -\omega t)$, with azimuthal mode number $l$ and
frequency $\omega /2\pi $, and (\textit{ii}) an ion at radial position $r$
with energy $\mathcal{E}$ and guiding center orbital angular momentum $%
L=m_{i}\Omega r^{2}$ with respect to the $z$ axis.
Each time a quantum of energy $\delta \mathcal{E}=$ $\hbar \omega $ is
absorbed by this ion, a quantum of angular momentum $\delta L=$ $\hbar l$ is
also absorbed. This angular momentum gain is then dissipated as a result of
the collisional torque $\Gamma $ Eq.~(\ref{torq}). We call $w_{l\omega }$
the steady state power, per particle, needed to sustain the angular momentum 
$L$ 
\begin{equation}
w_{l\omega }=\frac{d\mathcal{E}}{dt}=\frac{\omega }{l}\frac{dL}{dt}=\frac{%
\omega }{l}\Gamma =2q\nu _{i}\frac{\omega \Omega ^{3}B}{l{\omega _{i}}^{3}}%
r^{2}\text{,}  \label{ww1}
\end{equation}
where we have balanced the wave driven angular $dL/dt$ momentum gain by the
nonlinear Ohmic loss Eq.~(\ref{torq}). Starting from this single particle
power absorption $w_{l\omega }$, the efficiency of angular momentum
generation can be easily expressed.

Before expressing this efficiency we will recover this result Eq.~(\ref{ww1}%
) with the full picture of the Hamiltonian dynamics along the lines used in
the recent studies on wave driven rotational transform and transport driven
current generation in centrally fuelled discharges~\cite{Rax2017,Rax2018}.

Consider an electromagnetic wave displaying orbital angular momentum around $z$ and linear momentum along $z$, that is to say with typical space-time
periodicity of the type $f\left( r\right) \sin (l\theta +k_{\mathbf{\Vert }%
}z-\omega t)$ where $l$ is the azimuthal mode number and $k_{\mathbf{\Vert }%
} $ the parallel wave vector. Near a given point $r$ this cylindrical wave
can be viewed locally as a plane wave $\sin (k_{\bot }x+$ $k_{\mathbf{\Vert }%
}z-\omega t)$ where $\left( x,y,z\right) $ is a local set of Cartesian
coordinates centered on $r$ such that $y$ is directed along $\mathbf{e}_{r}$
, $x$ is directed along $\mathbf{e}_{\theta }$ and $k_{\bot }=l/r$ is the
local perpendicular wave vector. When this electromagnetic wave propagates and is absorbed in a rotating
plasma, a certain amount of (\textit{i}) energy, (\textit{ii}) linear
momentum along the $z$ axis and (\textit{iii}) angular momentum around the $%
z $ axis, are exchanged between the wave and the particles.

The energy $\mathcal{E}$, the parallel momentum $mv_{\Vert }$ along $z$ and
the cyclotron velocity $v_{c}$ around $\mathbf{B}$ of resonant particles
are changed through the wave interaction, but also the guiding center
positions $y_{g}$ across $\mathbf{B}$ as part of the momentum are no longer
free but bound to the static magnetic field through the invariance of the
canonical momentum along $y$~\cite{Rax2017,Fisch1992,Fisch1993}. This variation of the guiding
center positions $y_{g}$ drives a radial charge separation between magnetic
surfaces and provides orbital angular momentum deposition (rotation around $%
z $) inside the magnetized plasma column in addition to energy (heating) and
linear momentum (current generation along $z$) depositions~\cite{Fisch1978,Fisch1987}.

This picture of resonant wave-particle interaction has already been
presented in previous studies, both \ for plane waves and for cylindrical
waves~\cite{Rax2017,Rax2018,Fisch1992,Fisch1993}. It has been demonstrated, on the basis of Hamilton's
equations, that the increments of the various dynamical variables $\mathcal{E%
}$, $v_{\Vert }$, $v_{c}$ and $y_{g}$ are not independent, but fulfill
\begin{equation}
\frac{dy_{g}}{d\mathcal{E}}=\frac{k_{\bot }}{m_{i}\omega _{i}\omega }\text{
, }\frac{dv_{c}}{d\mathcal{E}}=\frac{\omega _{i}}{m_{i}v_{c}\omega }\text{, }%
\frac{dv_{\Vert }}{d\mathcal{E}}=\frac{k_{\Vert }}{m_{i}\omega }\text{.}
\end{equation}
During a small time $dt$ the steady state power transferred from the wave to
the particle is simply $d\mathcal{E}=w_{l\omega }dt$ \ and for continuous
absorption this power induces a radial particle velocity
\begin{equation}
\frac{dr}{dt}=\frac{dy_{g}}{d\mathcal{E}}\frac{\partial \mathcal{E}}{%
\partial t}=\frac{l}{qB\omega r}w_{l\omega }\text{,}  \label{dfgh}
\end{equation}
where we have used the relation between global cylindrical and local
plane waves $k_{\bot }=l/r$. This wave induced radial current is
continuously short circuited by the collisional velocity $\mathbf{w}_{i}$
given by Eq.~(\ref{dert}). At a radius $r$, the power needed to sustain the
RF driven G-J charge separation against collisional relaxation is locally
given by the balance $\mathbf{w}_{i}=\mathbf{e}_{r}dr/dt$ between Eqs.~(%
\ref{dert}) and (\ref{dfgh}) 
\begin{equation}
2\nu _{i}\frac{\Omega ^{2}E}{{\omega _{i}}^{3}B}=\frac{lw_{l\omega }}{qB\omega
r}\text{.}
\end{equation}
Thus, as $\Omega =E/Br$, we recover is the result Eq.~(\ref{ww1}) obtained
on the basis of a simple quantum transition argument.

These results are more conveniently presented if we consider the total power 
$W_{l\omega }$ needed to sustain the full plasma column rotation. This
global power is proportional to the energy $U_{\Omega }$ stored in the
rotation and this relation can be expressed by the efficiency 
\begin{equation}
\frac{W_{l\omega }}{U_{\Omega }}=\frac{2\int nw_{l\omega }2\pi rdr}{\int
nm_{i}\Omega ^{2}r^{2}2\pi rdr}=\frac{4\omega }{l}\frac{\Omega \nu _{i}}{%
{\omega _{i}}^{2}}\text{.}  \label{eff}
\end{equation}
Equation (\ref{eff}) is one of the main new result of this paper. It
expresses the efficiency of angular momentum generation when the wave $%
\left( \omega ,l\right) $ sustains a rigid body rotation $\Omega $ in a
magnetized fully ionized plasma $\left( \nu _{i},\omega _{i}\right) $.

According to Eq.~(\ref{ww1}), to sustain a rigid body rotation in a uniform
plasma the power deposition profile $w_{l\omega }\left( r\right) $ must be
quadratic with respect to the radius $r$. To sustain a
rigid body rotation in an axisymmetric configuration such as the one
depicted in Fig.~\ref{Fig:Fig3}, the power deposition profile $w_{l\omega }\left( \Psi
\right) $ must follow the scaling of the ratio $\nu _{i}r^{2}{B_{z}}^{2}/B^{2}$
as a function of $\Psi $.

Note that this analysis of the angular momentum generation process is simpler than the wave driven current generation because the relaxation does
not involve a kinetic description of the resonant particles~\cite{Fisch1978,Fisch1987}. Here
the short circuiting of the wave driven charge separation is ensured by a
bulk inertial ionic current Eq.~(\ref{curent}) even if the wave power is
absorbed by a minority population.

Eq.~(\ref{eff}) is a new universal result independent of the particular wave
branch and wave-particle resonance. The relation Eq.~(\ref{eff}) indicates
that low frequency waves with high azimuthal number are more efficient to
drive rotation through orbital angular momentum deposition.

\section{Finite Larmor radius and ion-ion dissipations}
\label{Sec:Sec7}

In the previous sections we assumed that the temperature difference $%
k_{B}\Delta T$ between two magnetic surfaces $\Psi _{1}$ and $\Psi _{2}$ is
smaller than the voltage drop $q\Delta \phi $ between these two surfaces,
thus the pressure force can be neglected in front of the electric force as $%
\left| \mathbf{\nabla }nk_{B}T\right| \ll nq\left| \mathbf{\nabla }\phi
\right| $. This ordering is fulfilled by the fast rotating discharges of
interest for thermonuclear fusion and allows to study separately \textit{%
inertial effects} and \textit{finite Larmor radius diamagnetic effects}
which are smaller. Besides diamagnetic effects, ion-ion Coulomb collisions
drive a friction force such that an additional dissipation takes place if $%
d\Omega /d\Psi \neq 0$ as the fast magnetic surfaces will transfer angular
momentum to the slow ones through ion-ion friction. This effect can be
minimized down to a very small value since a well shaped power deposition
profile $w_{l\omega }\left( \Psi \right) $ can be identified from the
relation Eq.~(\ref{ww1}) in order to ensure $d\Omega /d\Psi $ = $d^{2}\phi
/d\Psi ^{2}$ $\sim $ $0$.

In this section we set up the frame to identify the basic scaling and
ordering of thermal and viscous effects providing collisional relaxations of
G-J charges. A deeper analysis of the impact of these finite Larmor radius
effects and ion-ion collision effects is left to a future work.

Consider a fully ionized axisymmetric plasma\ such that the ion pressure is
equal to the electron one everywhere and the total pressure is $P\left( \Psi
\right) $. The ion and the electron diamagnetic velocities, $\mathbf{v}%
_{i}^{*}$ and $\mathbf{v}_{e}^{*}$, are azimuthal and they flow in opposite
directions, $\mathbf{v}_{i}^{*}\left( =-\mathbf{v}_{e}^{*}\right) $ can be
expressed as 
\begin{equation}
\mathbf{v}_{i}^{*}=\mathbf{b}\times \frac{\mathbf{\nabla }\Psi }{2nqB}\frac{%
dP}{d\Psi }\text{.}
\end{equation}
The collisional friction between the electron and ion flows is the source of
an azimuthal force $\mathbf{F}_{i}^{*}$ on the ions and $\mathbf{F}%
_{e}^{*}\left( =-\mathbf{F}_{i}^{*}\right) $ on the electrons 
\begin{equation}
\mathbf{F}_{i}^{*}=-2m_{i}\nu _{i}\mathbf{v}_{i}^{*}=-\frac{\nu _{i}}{\omega
_{i}}\frac{dP}{d\Psi }\mathbf{b}\times \frac{\mathbf{\nabla }\Psi }{n}\text{.%
}
\end{equation}
The factor $2$ comes from the relation $\mathbf{v}_{i}^{*}-\mathbf{v}%
_{e}^{*}=2\mathbf{v}_{i}^{*}$. These azimuthal friction forces are the
source of a cross-field ion flow $\mathbf{V}_{i}^{*}$ and a cross-field
electron flow $\mathbf{V}_{e}^{*}\left( =\mathbf{V}_{i}^{*}\right) $ 
\begin{equation}
\mathbf{V}_{i}^{*}=\frac{\mathbf{F}_{i}^{*}\times \mathbf{B}}{qB^{2}}=\frac{%
\nu _{i}}{nqB\omega _{i}}\frac{dP}{d\Psi }\mathbf{b}\times \left( \mathbf{b}%
\times \mathbf{\nabla }\Psi \right) \text{.}  \label{coco}
\end{equation}

Now we consider the Coriolis force in the (co)rotating frame. The two small
collisional flows $\mathbf{V}_{e}^{*}$ and $\mathbf{V}_{i}^{*}$ Eq.~(\ref
{coco}) are responsible for small azimuthal Coriolis forces $\mathbf{f}%
_{i}^{*}$ and $\mathbf{f}_{e}^{*}\left( \ll \mathbf{f}_{i}^{*}\right) $ 
\begin{equation}
\mathbf{f}_{i}^{*}=2m_{i}\Omega \mathbf{V}_{i}^{*}\times \mathbf{e}_{z}=2%
\frac{\nu _{i}\Omega }{n{\omega _{i}}^{2}}\frac{dP}{d\Psi }\left[ \mathbf{b}%
\times \left( \mathbf{b}\times \mathbf{\nabla }\Psi \right) \right] \times 
\mathbf{e}_{z}\text{.}
\end{equation}
This Coriolis collisional force drives a cross-field ion flow and a smaller
cross-field electron flow. The current associated with these flows is 
\begin{equation}
\mathbf{j}^{*}=n\frac{\mathbf{f}_{i}^{*}\times \mathbf{b}}{B}=2\frac{\nu
_{i}\Omega }{B{\omega _{i}}^{2}}\frac{dP}{d\Psi }\left( \left[ \mathbf{b}%
\times \left( \mathbf{b}\times \mathbf{\nabla }\Psi \right) \right] \times 
\mathbf{e}_{z}\right) \times \mathbf{b}\text{.}
\end{equation}
The direction of this flow is along the electric field as 
\begin{equation}
\left( \left[ \mathbf{b}\times \left( \mathbf{b}\times \mathbf{\nabla }\Psi
\right) \right] \times \mathbf{e}_{z}\right) \times \mathbf{b}=-\frac{%
\partial \Psi }{\partial r}\frac{\mathbf{E}}{E}\text{,}
\end{equation}
and its amplitude is proportional to the momentum exchange frequency. The
final diamagnetic current providing G-J charge relaxation is thus given by
\begin{equation}
\mathbf{j}^{*}=2\nu _{i}\frac{\Omega }{{\omega _{i}}^{2}B}\left| \frac{%
\partial P}{\partial r}\right| \frac{\mathbf{E}}{E}\text{.}  \label{diamag}
\end{equation}

The associated conduction is linear and the conductivity, independent of the
electric field, displays a fourth power scaling with respect to the magnetic
field. The finite Larmor radius linear conductivity associated with Eq.~(\ref
{diamag}) is smaller than the nonlinear inertial one described by Eq.~(\ref
{curent}) since 
\begin{equation}
\frac{j}{j^{*}}=\frac{nq}{E}\frac{\Omega ^{3}}{\omega _{i}}\left( \frac{%
\partial \Psi }{\partial r}\right) ^{2}/\frac{\partial P}{\partial r}\sim
\left( \frac{\Omega }{\omega _{i}}\frac{r}{\rho _{i}}\right) ^{2}\text{.}
\end{equation}

Besides this small diamagnetic effect, ion-ion collisions offer also a
possibility to relax the G-J charges. If we assume that the ion viscosity is
given by the Braginsky relation~\cite{Helander2002,Braginskii1965}, then the viscous ion-ion friction
force per particle is
\begin{equation}
\mathbf{F}_{i\rightarrow i}=\frac{3k_{B}T}{10\sqrt{2}}\frac{\nu }{{\omega
_{i}}^{2}}\Delta \mathbf{v}_{i}\text{.}
\end{equation}
This force cancels for the Brillouin flow, but for a Ferraro flow on each
magnetic surface the isorotation law $\Omega \left( r,z\right) =\Omega
\left( \Psi \right) $ is fulfilled and if $d\Omega /d\Psi \neq 0$ the fast
magnetic surfaces accelerate the slow ones and the slow magnetic surfaces
slow down the fast ones through ion-ion momentum transfer $\mathbf{F}%
_{i\rightarrow i}$. The azimuthal component of the force $\mathbf{F}%
_{i\rightarrow i}$ describing this momentum exchange is given by the
classical expression for the azimuthal viscous stress component
\begin{equation}
\mathbf{e}_{\theta }\cdot \mathbf{F}_{i\rightarrow i}=\frac{3k_{B}T}{10\sqrt{%
2}}\frac{\nu }{{\omega _{i}}^{2}}\left( \frac{1}{r^{2}}\frac{\partial }{%
\partial r}r^{3}\frac{\partial \Omega }{\partial r}+r\frac{\partial
^{2}\Omega }{\partial z^{2}}\right) \text{,}  \label{strees}
\end{equation}
where we have assumed flat density and temperature profiles. The azimuthal
component of the viscous force drives a cross-field flow $%
\mathbf{F}_{i\rightarrow i}\times \mathbf{B/}qB^{2}\mathbf{\ }$which is
colinear with the electric field $\mathbf{E}$ and the viscous current $%
\mathbf{j}_v=nq\mathbf{F}_{i\rightarrow i}\times \mathbf{B/}qB^{2}$ is given by 
\begin{equation}
\mathbf{j}_v=\frac{3nk_{B}T}{10\sqrt{2}}\frac{\nu B}{{\omega _{i}}^{2}E^{2}}%
\Omega ^{2}\left( \frac{\partial }{\partial r}r^{3}\frac{\partial \Omega }{%
\partial r}+r^{3}\frac{\partial ^{2}\Omega }{\partial z^{2}}\right) \frac{%
\mathbf{E}}{E}\text{.}  \label{viscous}
\end{equation}
This conduction process is linear with respect to the electric field but is 
\textit{non local} as it involves $\partial E/\partial r$, $\partial
^{2}E/\partial r^{2}$ and $\partial ^{2}E/\partial z^{2}$ through the
derivatives of $\Omega =E/Br$.

In order to set up an ordering between Ohmic dissipation and viscous damping
we simply have to compare the azimuthal Coriolis force $\mathbf{f}_{i}$ given
by Eq.~(\ref{cor}) with the azimuthal viscous stress component $\mathbf{e}%
_{\theta }\cdot \mathbf{F}_{i\rightarrow i}$ given by Eq.~(\ref{strees}).
Viscous damping is unimportant if $\mathbf{e}_{\theta }\cdot \mathbf{F}%
_{i\rightarrow i}<\mathbf{e}_{\theta }\cdot \mathbf{f}_{i}$, which rewrites
\begin{equation}
\left| \frac{1}{\Omega ^{3}r^{3}}\frac{\partial r^{3}\partial \Omega }{%
\partial r\partial r}\right| \mathbf{<}10\frac{\sqrt{m_{e}m_{i}}}{k_{B}T}%
\text{.}
\end{equation}
Note that this inequality is easily fulfilled: (\textit{i}) for rigid body
rotation $\Omega \left( r,z\right) =\Omega _{0}$, (\textit{ii}) for
Keplerian rotation $\Omega \left( r,z\right) =L/r^{2}$ under the assumption of homogeneous viscosity and (\textit{iii}) if
the power deposition profile $w_{l\omega }\left( \Psi \right) $ is tailored
such that $d\Omega /d\Psi $ $\sim $ $0$. For a generic isorotation $d\Omega
/d\Psi \neq 0$ we have to identify the critical radius where the opposite
ordering takes place. This deeper analysis of viscous damping is left to an
other study.

To summarize this section, we have addressed the issues of finite Larmor
radius and ion-ion collisions as candidate mechanisms to relax G-J charges
in rotating plasmas, both effect have been quantified and shown to be
smaller than the nonlinear Ohmic dissipation Eq.~(\ref{curent}) for typical
plasmas of interest such as Brillouin rotation and generic isorotation with $d\Omega /d\Psi $ $\sim $ $0$.

\section{Summary and conclusion}
\label{Sec:Sec8}

The main three new results of this study are given by Eqs.~(\ref{curent}), (%
\ref{qqq}) and (\ref{eff}) obtained in Secs.~\ref{Sec:Sec4}, \ref{Sec:Sec5} and \ref{Sec:Sec6}.

The relation Eq.~(\ref{curent}) provides the general expression of the Ohmic
nonlinear current short circuiting G-J charges, Eq.~(\ref{gjgl}), in an
axisymmetric rotating plasma described by the magnetic flux $\Psi \left(
r,z\right) $ and electric potential $\phi \left( \Psi \right) $. Compared to the classical Braginsky expression for the conductivity
perpendicular to the magnetic field, the axisymmetric rotating plasma displays a $B^{-6}$ scaling with
respect to the magnetic field strength instead of the the classical $B^{-2}$
scaling. This scaling and the small value of
the current for typical plasmas parameters explain the high RF efficiency
of rotation generation in fully ionized magnetically confined plasmas.

This efficiency, as well as the quality factor for energy storage $Q$, are
given by Eqs.~(\ref{qqq}) and (\ref{eff}). To keep these expressions simple, we have neglected the curvature $R^{-1}$ of the field lines through
the approximation $\partial \Psi /\partial r$ $=$ $rB_{z}\approx rB$ or $%
\partial \Psi /\partial r$ $=$ $E_{r}/\Omega \approx E/\Omega $. Taking into account the curvature leads to the expression of the
power associated with the Ohmic torque 
\begin{equation}
\iiint \mathbf{\Gamma }\cdot \mathbf{\Omega }d\tau =2q\iiint \nu _{i}\frac{%
\Omega ^{4}}{{\omega _{i}}^{3}}\frac{E_{r}^{2}}{E^{2}}Br^{2}d\tau \mathbf{e}%
_{z}\text{,}
\end{equation}
in place of Eq.~(\ref{pow}). The general relations, with full account of
finite curvature $R$, can be then easily derived on the basis of Eq.~(\ref{curent}%
). Whether or not curvature is accounted for, the quality factor and efficiency will still be written as the ratio of two integrals with geometrical factors $%
\partial \Psi /\partial r$ and will display the same final scaling as Eqs.~(%
\ref{qqq}) and (\ref{eff}) with respect to the plasma parameters.

The original result given by Eq.~(\ref{curent}) completes the results of a
previous study where the relaxation of a Brillouin flow in a weakly ionized
rotating plasma was analyzed on the basis of the exact solution of the
uncoupled fourth order algebraic equations describing electron and ion
vorticities~\cite{Rax2015}. This problem was much simpler than the present one since
the geometry was simpler and the electrons and ions dynamics were
uncoupled because neutral collisions, rather than Coulomb collisions, ensured
relaxation. The relation Eq.~(\ref{curent}) is directly relevant to the
analysis of the power balance in innovative magneto-electric toro\"{i}dal
traps in the limit of very large aspect ratio~\cite{Rax2017,Stix1970,Stix1971}.

Eq.~(\ref{eff}) gives the efficiency of wave orbital angular momentum
conversion into plasma orbital angular momentum. It shows that the conversion efficiency does not depend on the details of the wave dispersion and the wave-particle
resonance. This analysis completes the
results obtained on wave driven rotational transform where the problem of
poloidal rotation sustainment was addressed for the purpose of toroidal
magnetic confinement~\cite{Rax2017}. There the efficiency was directly compared to
classical current generation efficiency~\cite{Fisch1978,Fisch1987}, and was constrained from
the very beginning of the study by the goal of a large rotational transform
to counteract the vertical magnetic drift of the toroidal trap. The high
efficiency of angular momentum generation in a closed toroidal configuration
predicted in these earlier studies is thus confirmed and extended here to open axisymmetric
configurations of the mirror type. It is to be noted that because of finite
aspect ratio, viscous damping is more important for toroidal traps than for
the mirror type discharges considered here.

Although this study was restricted to hydrogen plasmas, the generalization to $%
Z\neq 1$ single species plasmas is straightforward. On the other hand, the
generalization to a multiple ion species plasma is less straightforward. Yet, this extension appears necessary given the practical interest of these discharges~\cite{Kolmes2018}. 

This work also provides an electrical engineering point of view on axisymmetric rotating plasmas. The magnetic surfaces $\Psi _{1}$, $\Psi _{2}$, $\Psi _{3}$...
in Fig.~\ref{Fig:Fig3} can be viewed as a set of nested conducting cylindrical shells. The equivalent
network associated with two neighbouring shells is described by a capacitor
with capacitance $\mathcal{C}=\varepsilon _{\bot }\varepsilon _{0}S/d$ \ and a
nonlinear resistance $\mathcal{R}=d/S\sigma _{NL}$: 
\begin{equation}
\mathcal{C}=\frac{{\omega_{pi}}^{2}}{{\omega_{i}}^{2}}\frac{\varepsilon
_{0}S}{d}\text{ , }\mathcal{R}=\frac{{\omega _{i}}^{4}}{{\omega _{pi}}^{2}\Omega
^{2}}\frac{d}{2\varepsilon _{0}\nu _{i}S}
\end{equation}
where $d$ is the radial gap between the two neighboring magnetic surfaces, $%
S $ their surfaces and we have used Eq.~(\ref{jjj}) to evaluate the
conductivity $\sigma _{NL}$. We recover the scaling of the relaxation time
for this equivalent elementary RC cell, $\mathcal{RC}={\omega _{i}}^{2}/2\nu
_{i}\Omega ^{2}$, already identified in Secs.~\ref{Sec:Sec2} and \ref{Sec:Sec5}. Thus we can set up
an operational transmission line model of axisymmetric discharge as a
radially distributed circuit with a well defined \textit{impedance per unit
length} along the radial direction: the \textit{inverse} \textit{capacitance
per unit length} is ${\omega_{i}}^{2}/\varepsilon _{0}S{\omega_{pi}}^{2}\ $%
\ and the \textit{resistance per unit length }is ${\omega
_{i}}^{4}/2\varepsilon _{0}\nu _{i}S{\omega _{pi}}^{2}\Omega ^{2}$. The RF
driven charge separation or the DC voltage drop sustainment, described in
Secs.~\ref{Sec:Sec5} and \ref{Sec:Sec6}, can be viewed as a continuous recharging of this distributed
capacity to compensate the short circuiting associated with the distributed
conductance.

Besides this electrical engineering model, the analysis and description of
the interplay between Coriolis, centrifugal and Coulomb friction forces,
offered in Sec.~\ref{Sec:Sec4}, provides a clear description of the G-J charge
relaxation process in fully ionized plasmas.

Finally, a promising feature identified in this study is the possibility to tailor the power deposition profile 
$w_{l\omega }\left( \Psi \right) $ in order to reach approximately a
Brillouin regime $d^{2}\phi /d\Psi ^{2}$ = $0$ in an axisymmetric
configuration $\Psi \left( r,z\right)$, that is to say to design a rotating plasma free of rotational shear. This is particularly interesting given that velocity shear is typically a
source of both instabilities and damping.


\section*{Acknowledgements}
The authors appreciate discussions with Mikhail
Mlodik. This work was supported, in part, by US DoE Contract No.
DE-SC0016072. One author(IEO) also acknowledges the support of DoE
Computational Science Graduate Fellowship (DoE Grant No. DE-FG02-97ER25308). 

\section*{References}
%

\end{document}